\documentclass[fleqn,usenatbib]{mnras}

\usepackage[T1]{fontenc}

\DeclareRobustCommand{\VAN}[3]{#2}
\let\VANthebibliography\thebibliography
\def\thebibliography{\DeclareRobustCommand{\VAN}[3]{##3}\VANthebibliography}

\usepackage{graphicx}	
\usepackage{amsmath}	
\usepackage{amssymb}	
\usepackage{newtxtext,newtxmath}

\title[The edge-on debris disc of GSC 07396-00759]{ALMA's view of the M-dwarf GSC\,07396-00759's edge-on debris disc: AU\,Mic's coeval twin}

\author[P. F. Cronin-Coltsmann et al.]{
\parbox{\textwidth}{Patrick F. Cronin-Coltsmann,$^{1,2}$\thanks{E-mail: patrick.cronin-coltsmann@warwick.ac.uk }
Grant M. Kennedy,$^{1,2}$
Christian Adam,$^{3}$
Quentin Kral,$^{4}$
Jean-Fran\c cois Lestrade,$^{5}$
Sebastian Marino,$^{6,7}$
Luca Matr\`{a},$^{8,9}$
Simon J. Murphy,$^{10}$
Johan Olofsson,$^{11,12}$
Mark C. Wyatt$^{7}$}
\\
\\
\parbox{\textwidth}{
$^{1}$Department of Physics, University of Warwick, Gibbet Hill Road, Coventry, CV4 7AL, UK\\
$^{2}$Centre for Exoplanets and Habitability, University of Warwick, Gibbet Hill Road, Coventry CV4 7AL, UK\\
$^{3}$Centro de Astronom\'ia (CITEVA), Universidad de Antofagasta, Avenida U. de Antofagasta, 02800, Antofagasta, Chile\\
$^{4}$LESIA, Observatoire de Paris, Universit\'e PSL, CNRS, Sorbonne Universit\'e, Univ. Paris Diderot, Sorbonne Paris Cit\'e, 5 place Jules Janssen, 92195 Meudon, France\\
$^{5}$LERMA, Observatoire de Paris, PSL Research University, CNRS, Sorbonne Universit\'es, UPMC Univ. Paris 06, 75014 Paris, France\\
$^{6}$Jesus College, University of Cambridge, Jesus Lane, Cambridge CB5 8BL, UK\\
$^{7}$Institute of Astronomy, University of Cambridge, Madingley Road, Cambridge CB3 OHA, UK\\
$^{8}$Centre for Astronomy, School of Physics, National University of Ireland Galway, University Road, Galway, Ireland\\
$^{9}$School of Physics, Trinity College Dublin, the University of Dublin, College Green, Dublin 2, Ireland \\
$^{10}$School of Science, The University of New South Wales, Canberra, ACT 2600, Australia\\
$^{11}$Instituto de F\'isica y Astronom\'ia, Facultad de Ciecias, Universidad de Valpara\'iso, Av. Gran Breta\~na 1111, Playa Ancha, Valpara\'iso, Chile\\
$^{12}$N\'ucleo Milenio Formaci\'on Planetaria - NPF, Universidad de Valpara\'iso, Av. Gran Breta\~na 1111, Playa Ancha, Valpara\'iso, Chile\\
}}

\date{Accepted XXX. Received YYY; in original form ZZZ}

\pubyear{2021}

\begin{document}
\label{firstpage}
\pagerange{\pageref{firstpage}--\pageref{lastpage}}
\maketitle

\begin{abstract}
We present new ALMA Band 7 observations of the edge-on debris disc around the M1V star GSC\,07396-00759. At $\sim20$\,Myr old and in the $\beta$ Pictoris Moving Group along with AU\,Mic, GSC\,07396-00759 joins it in the handful of low mass M-dwarf discs to be resolved in the sub-mm. With previous VLT/SPHERE scattered light observations we present a multi-wavelength view of the dust distribution within the system under the effects of stellar wind forces. We find the mm dust grains to be well described by a Gaussian torus at 70\,au with a FWHM of 48\,au and we do not detect the presence of CO in the system. Our ALMA model radius is significantly smaller than the radius derived from polarimetric scattered light observations, implying complex behaviour in the scattering phase function. The brightness asymmetry in the disc observed in scattered light is not recovered in the ALMA observations, implying that the physical mechanism only affects smaller grain sizes. High resolution follow-up observations of the system would allow investigation into its unique dust features as well as provide a true coeval comparison for its smaller sibling AU\,Mic, singularly well observed amongst M-dwarfs systems.
\end{abstract}

\begin{keywords}
circumstellar matter -- planetary systems -- stars: individual: GSC 07396-00759 -- submillimetre: planetary systems
\end{keywords}



\section{Introduction}
Many stars are host to discs of circumstellar matter. While the host star is still very young ($\lesssim$10 Myrs), these discs are composed of primordial dust and gas from the initial molecular cloud, and emit light in the near-infrared to millimetre ranges. As the discs age they lose their gaseous material to leave behind some dust and any planetesimals that have formed. The collisional grinding of these planetesimals produces cold secondary dust that is observed in the far-infrared to millimetre and is classified as a debris disc \citep[e.g.][]{Wyatt08,Hughes18}.
The dust produced by planetesimal collisions in a debris disc is constantly removed by radiation pressure as well as Poynting-Robertson drag and stellar wind forces, stellar wind forces being dominant over radiation pressure for grain removal around low-luminosity M-dwarfs \citep[e.g.][]{Wyatt99,Thebault08,Plavchan05,Augereau06,Reidemeister11}.
The defining observable features of a debris disc are typically a fractional luminosity  $L_{\rm{disc}}/L_{\star}$ of $\leq$10$^{-2}$, a lack of large amounts of warm dust emitting in the near-IR, and a lack of large quantities of H$_2$ gas. If any gas is present, e.g. CO, it is usually considered to be secondary, having also been released by planetesimal collisions \citep[e.g.][]{Marino16,Matra17,Kral19,Matra19}.

The Herschel DEBRIS survey detected debris discs around 17 percent of nearby main sequence FGK-type stars \citep{Sibthorpe18}, but found only 2 discs from 89 M-dwarfs \citep{Lestrade12,Kennedy13}.
However, a later Herschel survey of 21 planet-hosting late-type stars, of which 18 were M-dwarfs, with approximately twice the sensitivity to fractional luminosity as the DEBRIS survey detected 3 discs \citep{Kennedy18b}.
There is thus an open question \citep[e.g.][]{Plavchan05,Plavchan09,Gautier07,Heng13,Binks17,Luppe20} as to whether so few M-dwarf discs have been detected because they represent a fundamentally rarer and/or lower mass population to those of earlier type hosts, or whether the low luminosity of the host M-dwarfs, resulting in low disc fluxes and temperatures, hinders a similar population from being detectable.
Later type stars have a measured increase in planet occurrence rate \citep[e.g. ][]{Bonfils13,Dressing15,Mulders15}, which hints that perhaps efficient planet formation can affect the incipient debris disc by using up rocky material. Alternative scenarios for decreasing disc occurrence around late type stars include material stripping from stellar encounters \citep{Lestrade11},  photoevaporation of the primordial disc while the star is still present within its early cluster environment \citep{Adams04} and removal of dust by stellar-wind drag \citep{Plavchan09}. 
With so few known M-dwarf discs, it is important then to understand as fully as possible the discs that we do know and that we have well imaged. 

For a long period of time, the well imaged representative of M-dwarf discs has been the M1V star AU\,Microscopii. The excess of infrared radiation, the hallmark of circumstellar material, of AU\,Mic was first detected with IRAS \citep{Moshir90}. Only 9.72 $\pm$ 0.04\,pc \citep{Gaia18} distant and with a fractional luminosity of 4$\times 10^{-4}$ \citep{Matthews15}, AU\,Mic has been subject to detailed study ever since at a range of wavelengths observing both thermal emission and optical/Near-IR scattered light \citep[e.g.][]{Kalas04,Augereau06,Graham07,MacGregor13,Schneider14,Matthews15,Wang15,Wisniewski19}.

These high-resolution multi-wavelength views have resulted in myriad discoveries about the disc's physics. For example, \citet{Strubbe06} devise a 'birth ring' model for AU\,Mic where a parent population of planetesimals at 43\,au produces micrometer size dust grains that are then transported inwards by stellar wind drag and Poynting-Robertson drag, and outwards by radiation pressure and stellar wind ram pressure.
\citet{Boccaletti15,Boccaletti18,Grady20} observe fast moving dust features in scattered light travelling outwards along the disc, possibly dust 'avalanches' originating from the point of intersection of the birth ring and a second, inclined ring leftover from the catastrophic disruption of a large planetesimal \citep{Chiang17} or material released from a parent body on a Keplerian orbit closer to the star \citep{Sezestre17}.
\citet{Daley19} were able to estimate the sizes and masses of bodies within the disc through resolving its vertical structure.

These works highlight the value of obtaining resolved images in both thermal emission and scattered light. Fomalhaut\,C is as of yet the only other M-dwarf to have a fully resolved debris disc in thermal emission \citep{Coltsmann21}. However the disc was not detected with either HST/STIS nor VLT/SPHERE, the star is twenty times the age of AU\,Mic and the system may have a complicated disc-affecting dynamical history with its associated stars Fomalhaut\,A and B \citep{shannon14}. The complexity of the Fomalhaut system precludes Fomalhaut C from being a good representative.
With so much learned from the single system of AU\,Mic, it becomes increasingly valuable to have true coeval systems to compare AU\,Mic with and so that what we know of AU\,Mic's disc can be put into context.
 
As debris discs age, they deplete their reservoirs of planetesimals and are able to replenish less and less dust, meaning over time they become less bright \citep{Decin03,Rieke05}. In part, AU\,Mic owes its large fractional luminosity to its youth. 
AU\,Mic is a member of the $\beta$ Pictoris Moving Group (BPMG), a young \citep[$\sim$20\,Myr,][]{Bell15,Miret20} and nearby \citep[$\lesssim$100\,pc,][]{Shkolnik17} association of stars. \citet{Pawellek21} find a 75\% occurrence rate of discs around F type stars in the BPMG, a significantly higher rate than for field stars. 

An excellent place to search for AU\,Mic analogues is thus the BPMG, from which recently the late-type K7/M0 star CPD-72\,2713 \citep{Moor20,Tanner20} and the K1 star BD$+$45$^\circ$\,598 \citep{Hinkley21} have also recently had debris discs identified. 35 M-dwarfs in the BPMG have recently been observed in Band 7 with ALMA (Cronin-Coltsmann et al. in prep), yielding several new detections and one resolved disc. This resolved disc is GSC\,07396-00759, an M1V star at a distance of 71.4 pc. No WISE mid-IR excess is detected for GSC\,07396-00759, nor for AU\,Mic, making ALMA the best option for both detection and characterisation of such M-dwarf debris discs.
The disc of this star has been previously imaged in  near-IR scattered light with VLT/SPHERE IRDIS in both total intensity \cite[][hereafter S18; IRDIFS H2/H3]{sissa18} and polarimetric \cite[][hereafter A21; IRDIS DPI]{Adam21} modes, and is now detected for the first time in thermal emission. With a host star of similar spectral type and from the same young moving group, and therefore of very similar age, in addition to the disc being edge on, resolved in the sub-mm and well imaged in scattered light, GSC\,07396-00759's disc is a near perfect twin to AU\,Mic's, finally providing a coeval comparison.

S18's total intensity scattered light observations are subject to a strongly forward-scattering phase function that accentuates the brightness of the disc at small scattering angles and dims outer reaches off the major-axis. They find that the observed disc spine can be geometrically described up to 1.2\arcsec by an unflared disc of radius 96\,au and an inclination of 84.5$\pm$3.6$^{\circ}$ as they demonstrate in the lower panel of their Figure 2. They find a large brightness asymmetry, with the disc appearing brighter in the south-east by a factor $\sim$ 1.5-2. They also find ripples along the spine of the disc, and in the outer ranges of the disc they observe evidence of warp-like swept-back material, reminiscent of the swept-back 'wings' of HD\,61005 \citep[e.g.][]{Schneider14,Olofsson16}.

S18 then forward model the volumetric dust density distribution $n(r,z)$, as a double power law, see Equation \ref{eqn:doublepowerlaw}, and find that the disc density peaks at $r_0=$70 $\pm$1\,au and has a profile that is as expected from dust produced in a birth ring and pushed out by strong radial forces \citep{Strubbe06}.

A21 also observe disc emission extending to 1.3\arcsec (93\,au), as well as a moderate brightness asymmetry by a factor $\sim$ 1.4-1.6, and evidence of a warp in the disc on the north-west side. In contrast to S18, A21 model the disc as dust grains originating from a parent planetesimal belt at a radius $r_0$ with a Gaussian scale width $\delta_r$, equivalent to our Equation \ref{eqn:gaussian}. The dust grains then populate orbits defined by their interaction with stellar wind and radiation pressure forces before the scattering phase function is applied to derive the models to compare with the data. 

Through their modelling A21 find disc properties consistent with S18 except for a disc radius of 107$\pm$2\,au, but they find a degeneracy between their model radius and the anisotropic scattering factor $g$, which for higher values weights scattering efficiency to smaller scattering angles. A larger, more forward scattering $g$ diminishes flux at the ansae and focuses it at the disc centre, allowing for a larger model radius while still accurately describing the data. A21 do rerun their modelling with the disc radius fixed to the S18 result of 70\,au, and indeed a lower $g$ is then fitted, however the model residuals are noticeably poorer in the outer reaches of the disc: the lower radius model describes their data less well. 

A21 conclude that their new 107\,au estimate of the reference radius, i.e. the birth ring of planetesimals, is likely a better estimate than S18's 70\,au radius. Both S18's and A21's best-fitting model parameters can be found in Table \ref{tab:modelresults}.

Sub-mm observations trace larger dust grains that are less affected by pressure forces and retain their orbits closer to where they were produced, thus tracing more directly the location of the planetesimal birth ring. Resolved sub-mm observations such as those presented in this paper serve a key role in breaking the degeneracy between $g$ and $r_0$ and solving the discrepancy between the total intensity and polarimetric scattered light model radii. 

GSC\,07396-00759 is itself a wide separation companion of the well-studied close-binary V4046 Sgr at a distance of 12300\,au \citep{Torres06,Kastner11}. V4046 Sgr possesses both a gas-rich circumbinary disc and evidence of ongoing accretion \citep[e.g.][]{Stempels04,Oberg11,Rosenfeld13,Rapson15,Kastner18,Dorazi19,Brunner22}. The survival of the more primordial-like disc of V4046 Sgr may be attributed to its binary nature, as \citet{Alexander12} find close binaries to possess longer lived discs than single stars. Nevertheless, the association of the two systems draws into question the nature of the disc around GSC\,07396-00759 which the new ALMA observations presented herein can shed further light on.

This paper presents the new Band 7 ALMA observations of GSC\,07396-00759 in \S\ref{sec:obsv}, followed by a description of the modelling process and modelling results in \S\ref{sec:mod} and \S\ref{sec:Results}. In \S\ref{sec:Discuss} we present our analysis with respect to the previous scattered light observations and we place the disc in the context of both the growing M-dwarf disc population and the wider debris disc population across spectral types.

\section{ALMA Observations}\label{sec:obsv}

GSC\,07396-00759 was observed with ALMA in Band 7 (0.87\,mm, 345\,GHz) on April 6th 2018 under project 2017.1.01583.S as part of a larger survey of M-dwarfs in the $\beta$ Pictoris Moving Group (Cronin-Coltsmann et al. in prep). The observation used baselines ranging from 15 to 484\,m and 43 antennae. The average precipitable water vapour was $\sim$0.75\,mm. The total on source observing duration was 16 minutes. QSOs J1826-2924 and J1924-2914 were used for atmospheric and water vapour radiometer calibration;
J1826-2924 was used for phase calibration;
J1924-2914 was used for pointing, flux and bandpass calibration. 

The spectral setup comprised four windows centred on 347.937, 335.937, 334.042 and 346.042 GHz with bandwidth 2 GHZ and 128 channels for all but the last with width 1.875 GHz and 3840 channels. 
The last window was used to search for CO gas via the J=3-2 emission line, which has also been detected in another young debris disc around the M-dwarf TWA\,7 \citep{Matra19}.

The raw data were calibrated with the provided ALMA pipeline script in \textsc{casa} version 5.1.2-4 \citep{CASA}. To reduce the data volume the visibilities were averaged in 30 second intervals and down to two channels per spectral window for the continuum imaging. All images were generated with the \textsc{clean} algorithm in \textsc{casa}.

\subsection{Continuum Analysis}

\begin{figure}
        {\includegraphics[width=\columnwidth]
        {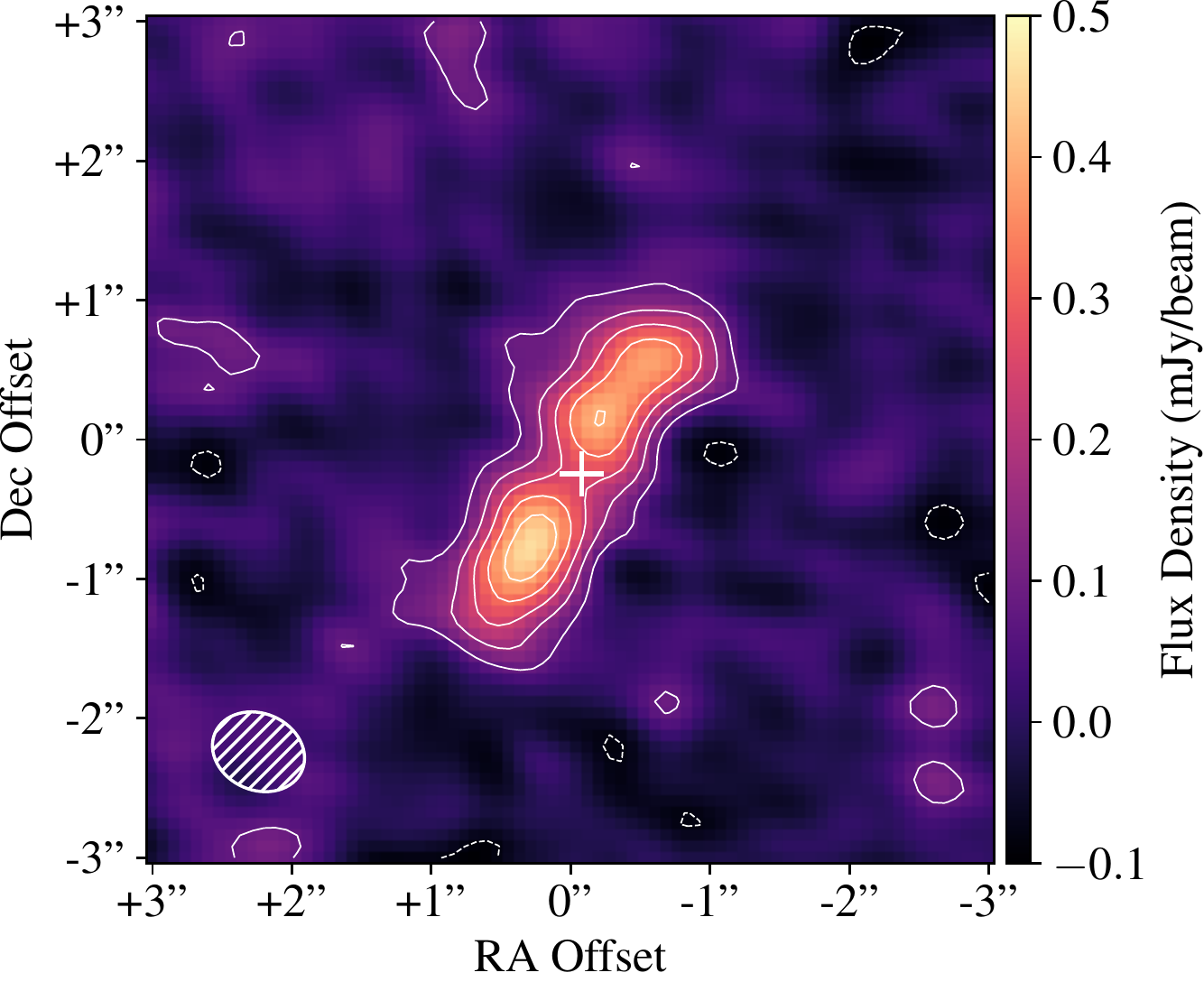}}
        \caption{\label{fig:Cont}  Naturally-weighted \textsc{clean} image of the disc around GSC\,07396-00759. The ellipse in the lower left corner shows the beam size of 0.68$\times$0.55\arcsec. The star is not detected. At a distance of 71.4\,pc the apparent disc radius is $\sim$70\,au. Contours are drawn at $\pm 2\sigma, 4\sigma, 6\sigma, 8\sigma, 10\sigma$ with $1\sigma = 40 \mu$Jy beam$^{-1}$. The Gaia location of the star is marked with a $+$ at 273$^\circ$35'31.2"$\pm 0.26$ mas $-$32$^\circ$46'11.09"$\pm 0.20$ mas. Zero offset is the ALMA image phase centre at 273$^\circ$35'31.3" $-$32$^\circ$46'10.9" (J2000).}
\end{figure}

Figure \ref{fig:Cont} shows a \textsc{clean} image of GSC\,07396-00759's disc. We use natural weightings for maximum signal to noise ratio (S/N). This weighting gives a synthesised beam with major and minor full width at half maxima (FWHM) of 0.68\arcsec (48.6\,au) and 0.55\arcsec (39.3\,au) respectively and a position angle (PA) of 66.4$^{\circ}$. We identify the standard deviation in an annulus exterior to the disc to be $\sigma = 40 \mu$Jy beam$^{-1}$. This noise is uniform throughout the central area where the disc is detected and the primary beam correction there is $< 10\%$. 

\begin{figure}
        {\includegraphics[width=\columnwidth]
        {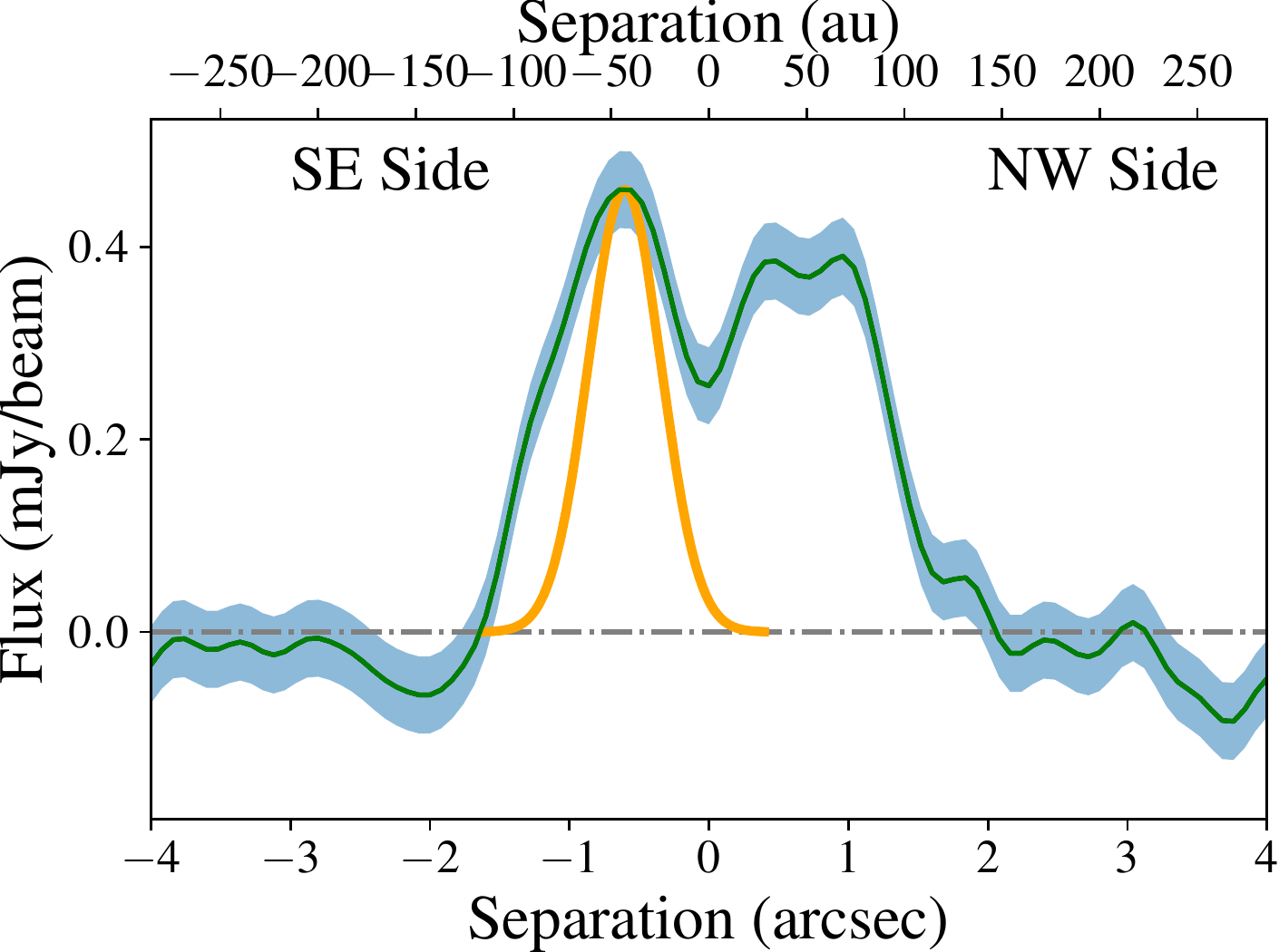}}
        \caption{\label{fig:RadialPlot} Profile of the disc along its major-axis; the flux of the centre pixel along the disc major-axis is plotted in green, blue swathes show the RMS and a Gaussian with the same FWHM as the beam is plotted in orange at the peak radial flux. Zero separation is the best fitting model centre from \S \ref{sec:mod}.}  
        
\end{figure}

The disc is continuously detected with at least 4$\sigma$ and detection peaks at 10$\sigma$ at the disc ansae. It is apparent that the flux constitutes a highly inclined ring with a radius of $\sim$1\arcsec and a position angle of $\sim-$30$^\circ$; the disc is unresolved along the minor-axis and the emission perpendicular to the major-axis appears to have a scale similar to the beam size, thus limiting the disc's maximum vertical extent to within 50\,au. The disc is radially resolved, as shown in Figure \ref{fig:RadialPlot}. The dip in the profile of the disc at $\sim$1\arcsec on the north-west side is on the scale of the beam, and therefore is likely the result of noise. As the south-east side only peaks 1$\sigma$ higher than the north-west, resulting in a difference in integrated flux of $\sim10\%$ with integrated fluxes of $\sim0.88$\,mJy and $\sim0.80$\,mJy respectively, we conclude that there is not strong evidence of asymmetry, but note that it was the south-east side that was significantly brighter in the scattered light observations of S18 and A21.

\section{Modelling}\label{sec:mod}

To extract probability distributions of the disc parameters we fit models directly to the u-v ALMA data. This is done by first creating a 3-dimensional disc model. A rotation from sky coordinates to model coordinates is calculated and used to find the corresponding model coordinate for each pixel in a volume centred on the star, and the given parameters are consulted to identify the model flux at each pixel location. This model disc is then collapsed into the sky plane in order to create a 2-dimensional image\footnote{https://github.com/drgmk/alma}. We use the \textsc{galario} package \citep{galario} to Fourier transform this image and to sample the u-v locations of the data to calculate a $\chi^2$. Posterior probability distributions of the model parameters are explored with the \textsc{emcee} package \citep{emcee}, an implementation of the Markov Chain Monte Carlo method in Python. We initiate our models near the solutions of previous test runs. We use 3000 steps and discard the first 2700 as the maximum auto-correlation length of the parameter chains is 270 steps. We use 100 walkers and verify that all the chains have converged upon completion.

We first implement a simple \textbf{Gaussian} torus model, with parameters defined by the following equation for a model dust volume density distribution $n(r,z)$:

\begin{equation}
\label{eqn:gaussian}
    n(r,z)\propto e^{-\frac{1}{2}\left(\frac{r-r_0}{\sigma_r}\right)^2}
    \times e^{-z^2/2h^2}
\end{equation}

\noindent where $r_0$ is the radius of peak flux, $\sigma_r$ is the Gaussian scale width (where the FWHM is 2.355*$\sigma_r$) and $h = r\times tan(\psi)$ and where $\psi$ is disc opening angle. 
In their scattered light modelling S18 use a fixed $\psi$ of 0.04, found by \citet{Thebault09} to be the 'minimum natural observed aspect ratio' for dust grains observed at visible to mid-IR wavelengths.
Dust grains observed at mm wavelengths whose orbits are not affected by radial forces are not expected to conform to this minimum aspect ratio, however trialling a similar model with $\psi$ as a free parameter finds $\psi$ unresolved in the ALMA data with a 3$\sigma$ upper limit of 0.18. We thus choose to fix $\psi$ to 0.04 in our  modelling for consistency with the modelling of S18.

Our second model follows the equation used in S18 for direct comparison, i.e. a \textbf{Double Power Law} with different density slopes interior and exterior to $r_0$:

\begin{equation}
\label{eqn:doublepowerlaw}
    n(r,z)\propto\left[\left(\frac{r}{r_0}\right)^{-2a_{\rm{in}}}+\left(\frac{r}{r_0}\right)^{-2a_{\rm{out}}}\right]^{-1/2}
    \times e^{-z^2/2h^2}
\end{equation}
where $a_{\rm{in}}$ and $a_{\rm{out}}$ are the inner and outer slopes respectively. 

Both models possess a parameter for the total flux of the disc as well as an inclination, a position angle and RA and Declination (Dec) offsets of the disc model centre from the ALMA phase-centre. The ALMA phase-centre is found to be slightly offset itself from the Gaia DR2 \citep{Gaia16,Gaia18} location of the star at the time of observation. All subsequent reported offsets have had that Gaia to phase-centre offset subtracted such that the offset measurements are relative to the Gaia location at the time of observation.
We also include a parameter for scaling the weightings of the u-v data points as their absolute uncertainty can be offset as described in \citet{Matra19b} and \citet{Kennedy20}.

We calculate a na\"ive 'plane-of-sky' eccentricity of the disc by simply taking the offset vector in the plane of the sky and dividing by the disc radius, and use the given offset and radius posteriors to form a posterior distribution of eccentricities with uncertainty reported that includes the uncertainty of ALMA's astrometric precision (calculated per \S10.5.2 of the ALMA Cycle 6 Technical Handbook\footnote{https://almascience.nrao.edu/documents-and-tools/cycle6/alma-technical-handbook}, to give 0.036\arcsec), the Gaia astrometric precision is negligible in comparison at 0.33 mas. As the disc is highly inclined, we cannot accurately discern any offset perpendicular to the plane of the sky. Because the true eccentricity, $e$, could be larger if the pericenter were not 90$^\circ$ to the line of sight (i.e. if $\omega$, the argument of pericentre, is not equal to 0$^\circ$), we do not quote an eccentricity measurement in Table \ref{tab:modelresults}, and we instead provide a 3$\sigma$ upper limit on $e\cos(\omega)$, i.e. the projection of the eccentricity vector along the major-axis of the disc. This upper limit is derived from the offset posterior distribution after the dot product is taken between the offset vector and the unit vector of the major-axis of the disc. 

\section{Results and analysis}\label{sec:Results}
\begin{table*}
\centering
\renewcommand{\arraystretch}{1.5}
\caption{Median disc parameters, $\Delta \chi ^2$ and $\Delta$BIC values for Gaussian and double power law models. Best fitting parameters for S18's total intensity scattered light modelling and A21's polarimetric scattered light modelling are also included for comparison. Uncertainties are the 16th and 84th percentiles. Offsets are measured from the disc model centre to the Gaia DR2 location of the star at the time of observation. Upper limits are at 3$\sigma$ above the mean, i.e. the 99.87th percentile. The disc flux uncertainty includes the 10\% ALMA flux calibration uncertainty. The $e\cos(\omega)$ upper limit includes the ALMA astrometric precision. The $e\cos(\omega)$ upper limit is calculated from the projection of the offset vector along the major-axis of the disc. $\Delta \chi ^2$ and $\Delta$BIC values relative to Gaussian model with values 721281.0 and 721384.6 respectively, calculated from a model produced using the median parameters.}

\label{tab:modelresults}
\begin{tabular}{lccccc}
\hline
Parameter & Gaussian & Double Power Law & Total Intensity Scattered Light & Polarimetric Scattered Light\\
\hline                                     
RA Offset (\arcsec)                 & $0.06^{+0.03}_{-0.03} $   & $0.06^{+0.03}_{-0.03}$       & $-$  & $-$  \\
Dec Offset (\arcsec)                & $0.00^{+0.04}_{-0.04} $   & $0.00^{+0.04}_{-0.03}$       & $-$  & $-$ \\
$e\cos(\omega)$ $3\sigma$ Upper Limit& $0.17$                 &   $0.16$                      & $-$ & $-$ \\
Inclination ($^{\circ}$)      & $85^{+3}_{-3}$       & $85^{+3}_{-3}$    & $82.7^{+0.1}_{-0.1}$ & $84.3^{0.3}_{-0.3}$  \\
PA ($^{\circ}$)               & $-32^{+1}_{-1}$         & $-32^{+1}_{-1}$    & $-31.1^{+0.1}_{-0.1}$& $-31.3^{+0.7}_{-0.7}$ \\
Disc Flux (mJy)               & $1.84^{+0.22}_{-0.21}$     & $1.83^{+0.21}_{-0.21}$    & $-$   & $-$        \\
Radius (au)                    & $70.2^{+4.1}_{-4.7}$    & $77.7^{+8.0}_{-7.1}$     & $69.9^{+0.9}_{-0.8}$& $107^{+2}_{-2}$  \\
Scale Width (au)               & $20.3^{+4.3}_{-4.1}$   & $-$                       & $-$      & $27^{+1}_{-1}$ \\
$\alpha_{in}$                   &  $-$  &    $3.5^{+4.5}_{-1.9}$    &   $2.8^{+0.2}_{-0.2}$    & $-$  \\
$\alpha_{out}$                  &  $-$  &    $-8.9^{+1.9}_{-2.0}$    &  $-2.6^{+0.1}_{-0.1}$     & $-$   \\
$N_{\text{Parameters}}$       & $8$                      & $9$                          &     $-$ & $-$   \\  
$\Delta\chi^2$                & $0$                      &  $+1.9$                     &    $-$   & $-$  \\ 
$\Delta$BIC                   & $0$                      & $+14.9$               &  $-$        & $-$  \\
   
\hline
\end{tabular}

\end{table*}

\begin{figure*}
        \centering
        \includegraphics[width=0.484\linewidth,height=0.468\linewidth]{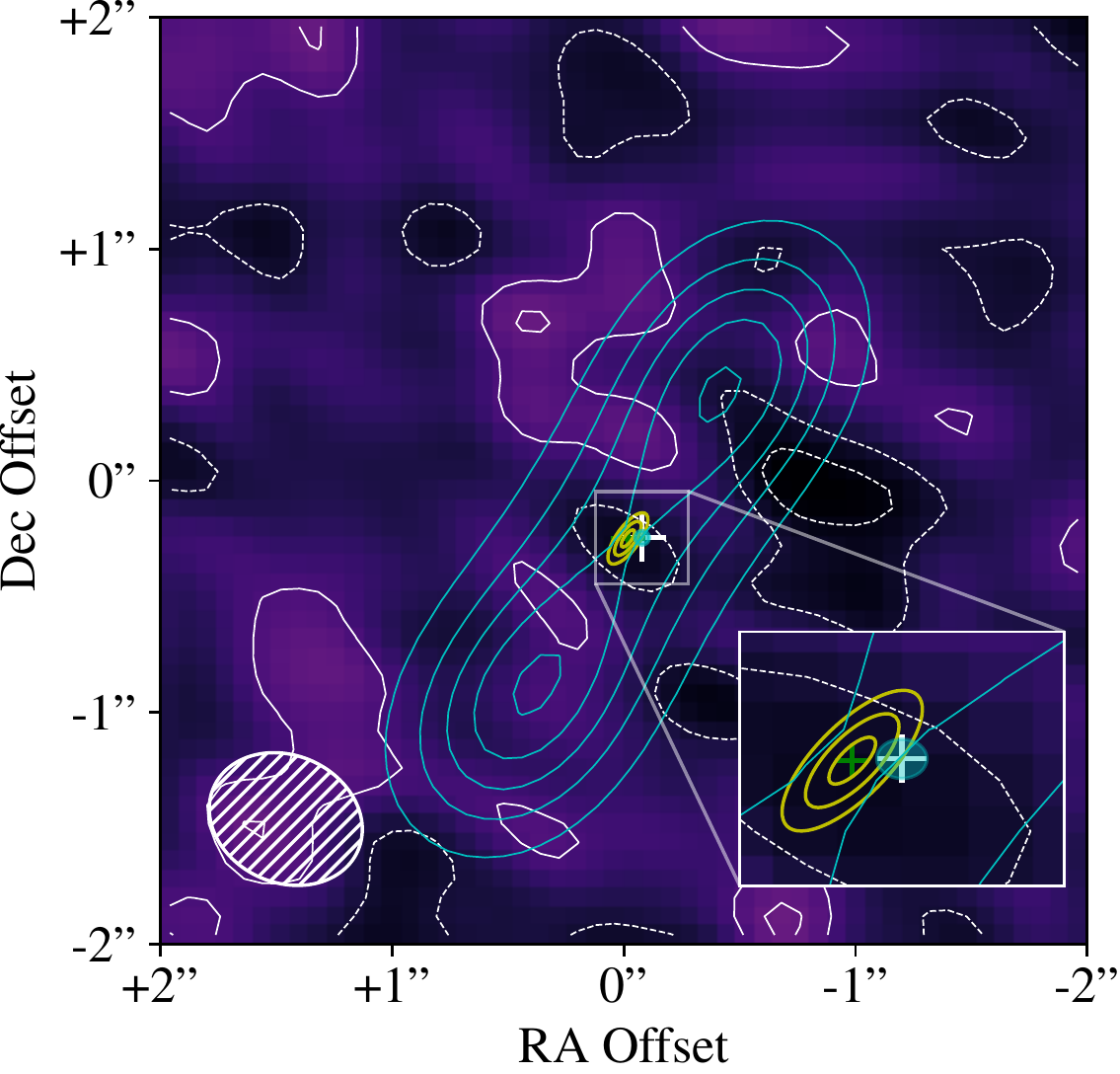}
        \hfill
        \includegraphics[width=0.51\linewidth,height=0.468\linewidth]{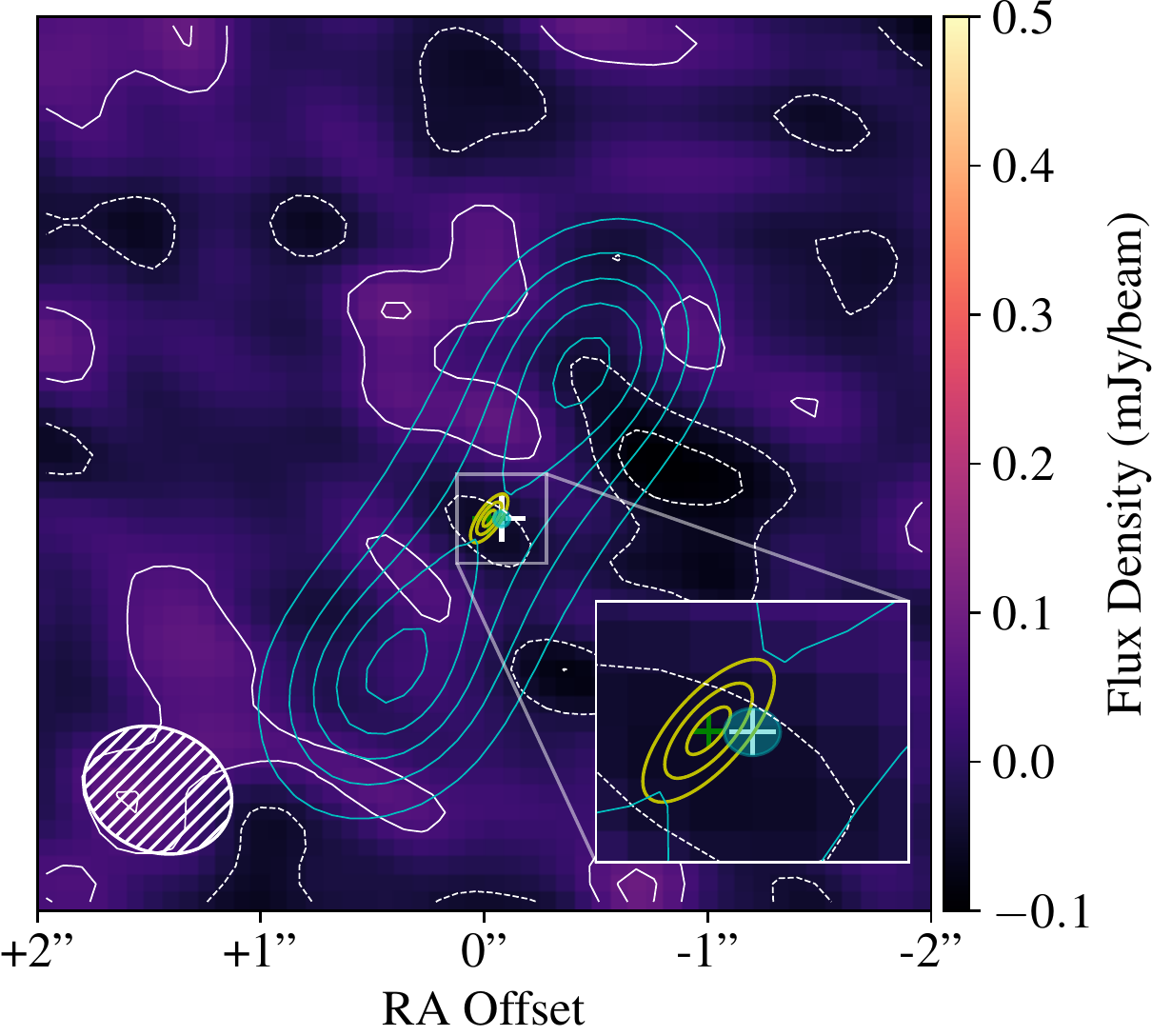}
    \caption{\label{fig:res}Naturally-weighted dirty images of the residuals after subtracting the individual models. Left: Gaussian model; Right: double power law model. Cyan contours show the models at $2\sigma, 4\sigma, 6\sigma, 8\sigma, 10\sigma$ and white contours show the residuals at $-1\sigma, -2\sigma, 1\sigma, 2\sigma$. The location of the star is marked with a $+$. Zero offset is the ALMA image phase centre at 273$^\circ$35'31.3" $-$32$^\circ$46'10.9" (J2000). The inset shows a zoom near the star to illustrate that the disc model centre uncertainty (yellow contours at $1\sigma, 2\sigma, 3\sigma$) and the ALMA astrometric precision (grey circle centred on the stellar location, $1\sigma$) are large enough that an offset is not significantly detected.}
\end{figure*}
\subsection{Gaussian torus model}\label{subsec:Torus}

This simple model serves as the default hypothesis, tracing an azimuthally symmetric parent planetesimal belt or 'birth ring' localised to one radius and with a radially symmetric dust distribution about that radius. 
Given the spatial resolution, the use of a Gaussian is not specific, equally a top-hat distribution or single-power law could have been used \citep[e.g.][]{Kennedy18b}, the important factors being radial symmetry and a measure of disc width.
The best-fitting parameters are presented in Table \ref{tab:modelresults} along with those of the double power law model and S18 and A21's scattered light models. The corner plot derived from the modelling is presented in Figure \ref{fig:CornerGau}. We show a dirty image of the residuals after subtraction of a model formed from the medians of the posterior-parameter distribution, with the model contours overplotted, in Figure \ref{fig:res} left. Figure \ref{fig:res} left also contains an inset showing the distribution of model disc centres in comparison to the stellar location and ALMA astrometric precision.

The residuals show no remaining structure, demonstrating that an azimuthally and radially symmetric model fits the data well. Only a single 2$\sigma$ residual overlaps with the bounds of the disc, a negative residual in the north-west. At 1-2$\sigma$ this feature is likely noise and accounts for the shape of the disc profile, Figure \ref{fig:RadialPlot}.

We use the median-parameter model to calculate a $\chi^2$ value of 721281.0 as well as calculating the Bayesian Information Criterion \citep[BIC;][]{Schwarz78} of 721384.5. The BIC penalises models for including extra parameters to identify whether an improvement in $\chi^2$ is justification to conclude a model is a significantly better fit to the data. It is defined as $BIC = \chi^2 + N_{\rm Parameters}\times\ln{N_{\rm dof}}$, and as we fit to a large number of visibilities ($N_{\rm dof} = 2\times N_{\rm vis} =2 \times 209496$) the inclusion of even a single extra parameter imposes a large penalty. A $\Delta$BIC of greater than 6 is considered 'strong' evidence and a $\Delta$BIC of greater than 10 is considered 'decisive' evidence that the lower valued model is significantly preferred to fit the data \citep{Kass95}.

The median parameters of this disc model largely align well with the parameters found from S18's and A21's scattered light modelling. The model inclination and position angle are within 1$\sigma$ of both. Our radius measurement is in significant agreement with the S18 model and significant disagreement with the A21 model. Our limit on the disc's Gaussian scale width is found to be consistent with A21's measurement.

We measure a median offset from the stellar position to the disc model centre of $5.0^{+2.1}_{-1.7}$\,au (before combination with the $\sim2.6$\,au ALMA astrometric uncertainty), visualised in the inset of Figure \ref{fig:res} left, this results in a median eccentricity of $0.08^{+0.05}_{-0.04}$ (after combination with the ALMA astrometric uncertainty). The distribution of offsets is more constrained perpendicular to the major-axis of the inclined disc as smaller shifts in offset in this direction will move comparatively more flux out of the bounds of the disc.
Given our modelling uncertainty and the sizeable ALMA astrometric uncertainty we conclude that this measurement is not significant evidence of underlying eccentricity. If an offset is present at this magnitude, a higher resolution observation with smaller parameter uncertainty and a smaller pointing uncertainty would be able to make a significant measurement.
We can instead place a 3$\sigma$ upper limit on eccentricity along the major-axis of the disc, $e\cos(\omega)$, of 0.17.

At this wavelength (0.87\,mm) we measure the flux of the disc to be $1.84^{+0.22}_{-0.21}$\,mJy (the uncertainties of which have been combined in quadrature with the 10\% ALMA flux calibration uncertainty), which informs our SED modelling in \S \ref{subsec:SED}.

\subsection{Double power law model}

This model serves as a direct comparison to S18's scattered-light model, to investigate whether the distribution of millimetre sized dust grains, visible in sub-mm thermal emission, overlaps with, or is similar in shape to, the distribution of micrometre sized dust grains, visible in the scattered light. 

Figure \ref{fig:res} right displays the dirty residual image for this model. It is immediately apparent that the two models produce nearly indistinguishable residuals, and this is attested to in the broad similarities of the parameter distributions. The only significant departure between the two models is the larger radius of the double power law model, although this comes with almost twice the uncertainty in values and remains consistent with the previous measurement. The median eccentricity of $0.07^{+0.03}_{-0.03}$ and the $e\cos(\omega)$ upper limit of 0.16 are both slightly smaller than the Gaussian's model, but not significantly. 

A steep outer slope of $\alpha_{\rm{out}}\sim-9$, which is incongruous with S18's scattered light model, is found for this model which necessitates a larger radius to still account for the flux most distant from the centre. The Gaussian distribution is comparatively wider and so can account for this flux without extending the radius, at the cost of a slightly increased concentration of flux at the centre. A shallower outer slope would increase the flux at this distant range, but would also necessarily increase the flux beyond it, which would be inconsistent with the data. This steep outer slope demonstrates a lack of evidence for millimetre dust grains beyond a confined birth ring, i.e. the radial forces of the system are too weak to transport the millimetre dust grains as they have for the micrometre dust grains. 

The inner slope measurement is more similar to the scattered light model, but has a very large uncertainty and so is not significant evidence of physical similarity. 

The double power law model is largely degenerate as shown in the modelling corner plot in Figure \ref{fig:CornerPow}. Radius is degenerate with both $\alpha_{\rm{in}}$ and $\alpha_{\rm{out}}$, and $\alpha_{\rm{in}}$ and $\alpha_{\rm{out}}$ are degenerate with each other: a smaller radius with steeper $\alpha_{\rm{in}}$ and shallower $\alpha_{\rm{out}}$ fits the data as well as a larger radius with shallower $\alpha_{\rm{in}}$ and steeper $\alpha_{\rm{out}}$.

This model is found to have both a larger $\chi^2$ and a significantly larger BIC, these measures together with the significant degeneracies of the model, the large uncertainties in the unique parameters $\alpha_{\rm{in}}$ and $\alpha_{\rm{out}}$, and the similarity of the residual images, allow us to conclude that the model is unnecessarily complicated given the data.

\subsection{Flux density distribution and fractional luminosity modelling}\label{subsec:SED}

\begin{figure}
        {\includegraphics[width=\columnwidth]
        {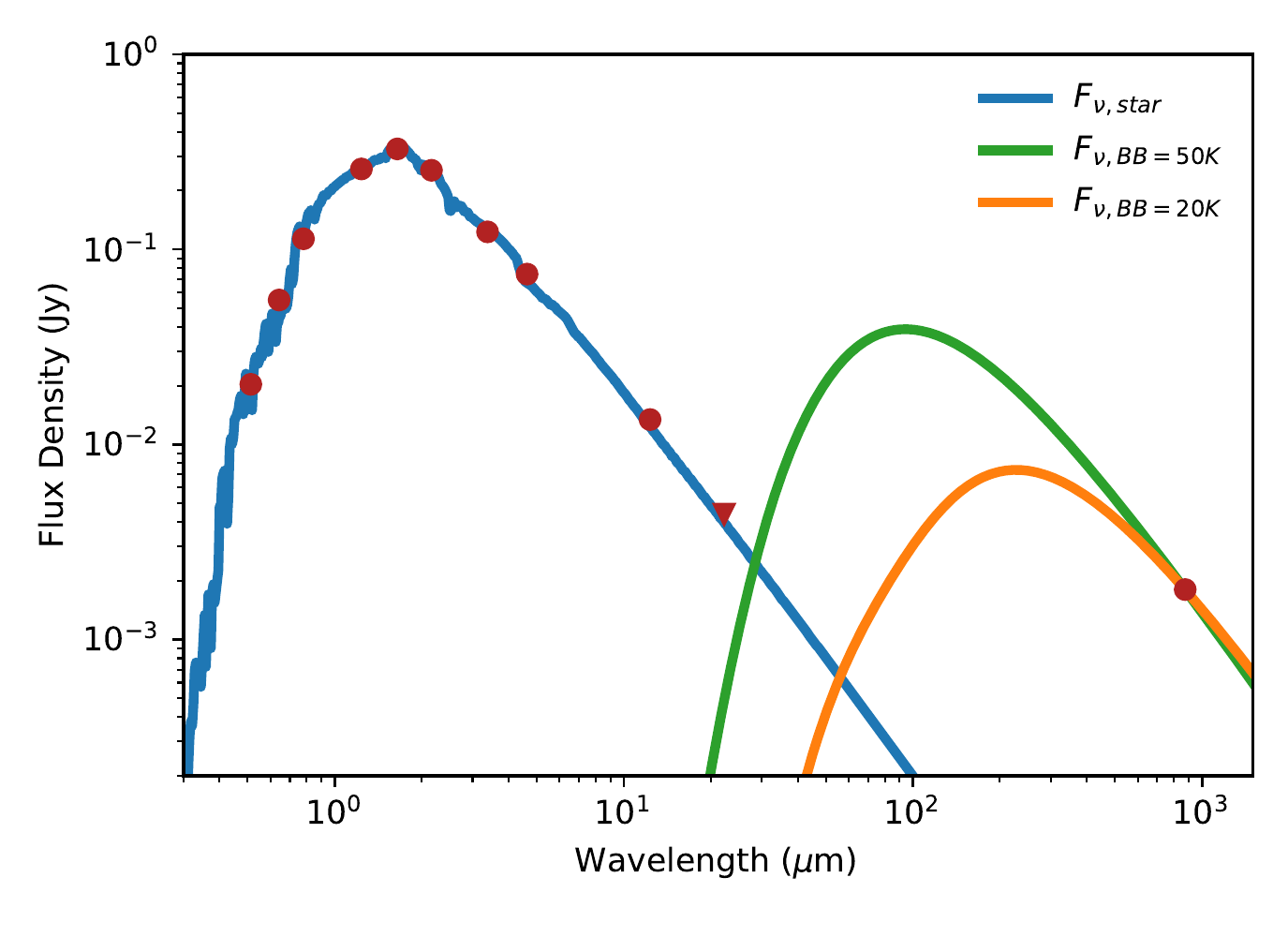}}
    \caption{\label{fig:sed}Example flux density distribution (SED) for the disc of GSC\,07396-00759. Dots are measured fluxes and triangles are 3$\sigma$ upper limits. The stellar photosphere model is in blue and example blackbody distributions at 20 and 50 K are fitted through the ALMA flux in orange and green respectively. With only one flux point measuring the thermal emission of the disc, a large range of temperatures and fractional luminosities could describe the disc.}
\end{figure}

\begin{figure}
        {\includegraphics[width=\columnwidth]
        {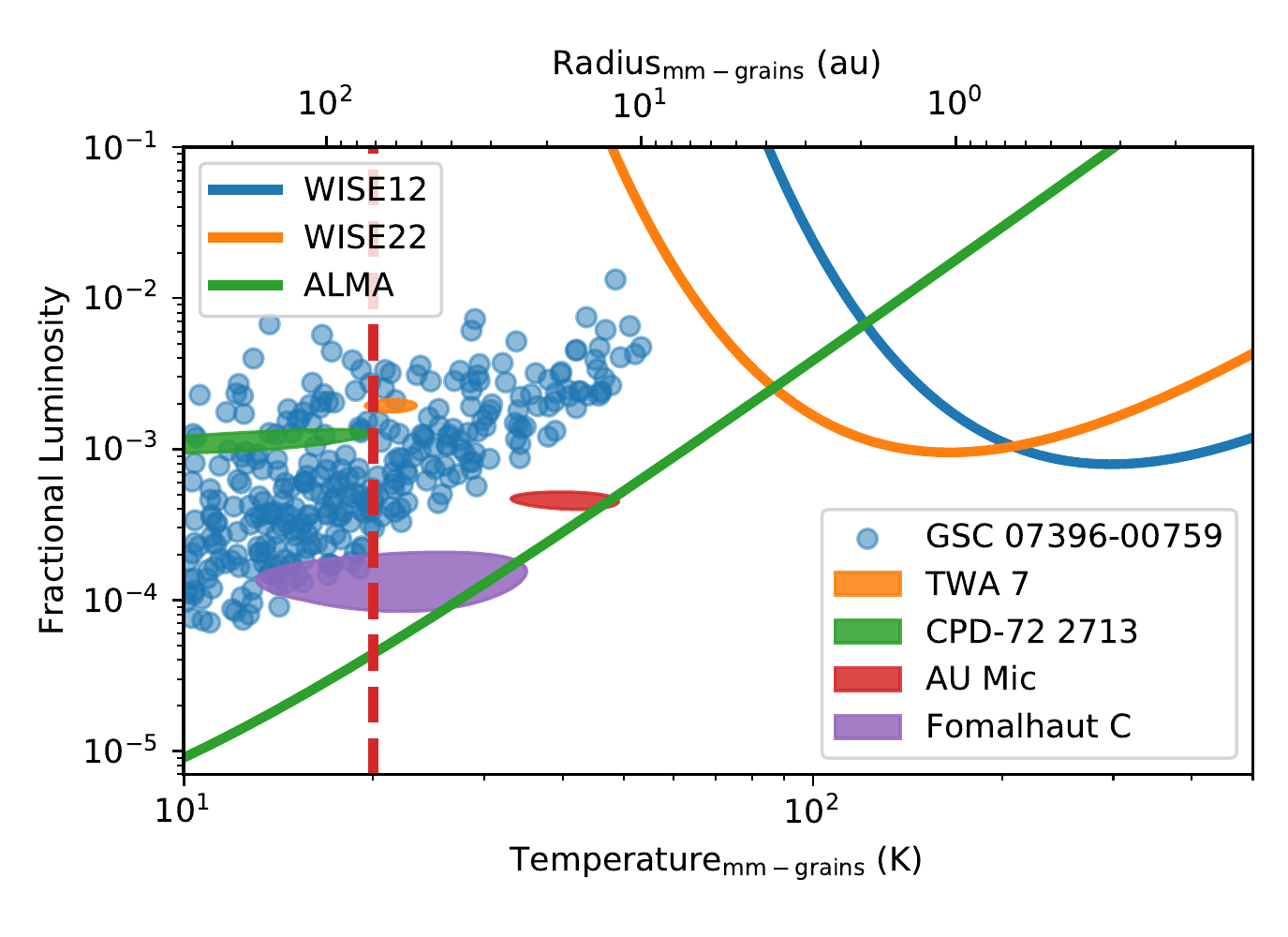}}
    \caption{\label{fig:ft}Plot of fractional luminosity against representative temperature/blackbody radius, i.e. the temperature and stellocentric radius of mm grains. Blackbody radius depends on host stellar temperature and is thus only accurate for GSC\,07396-00759. A selection of allowed models for the disc of GSC\,07396-00759 are plotted as blue circles. The distributions up to 3$\sigma$ following the same SED fitting procedure are shown for a selection of low mass host debris discs as coloured ellipses. The detection limits for the WISE 12 micron band, WISE 22 micron band and ALMA Band 7 are plotted as blue, orange and green curves respectively. The vertical red dashed line is placed at 70.2\,au, our best-fitting radius for GSC\,07396-00759's disc.}
\end{figure}

In addition to the ALMA observations reported here, GSC\,07396-00759 has been observed with Gaia \citep{Gaia16,Gaia18}, 2MASS \citep{2MASS06}, and WISE \citep{WISE10}, which we use to constrain the properties of the stellar photosphere and disc using the Spectral Energy Distribution (SED) fitting method described by \citet{Yelverton19}.

However, without spectral data points in the far-infrared from instruments such as Spitzer/MIPS or Herschel/PACS/SPIRE, the disc SED is poorly constrained. Figure \ref{fig:sed} demonstrates this problem with an example 50K blackbody distribution plotted through the measured ALMA flux; with only a single data point for the disc's thermal emission, in the sub-mm at the tail of the distribution, a large range of disc temperatures and luminosities can be fitted. Some constraints are still possible, as a dust temperature significantly greater than 50K would result in greater mid-infrared emission than is observed with WISE.

To provide a stronger constraint on fractional luminosity and temperature we model the disc emission with a more physical model using realistic grain optical properties and a size distribution \citep[e.g.][]{Augereau99,Wyatt02}. Here we assume astronomical silicates \citep{draine84}, though our results are fairly insensitive to the choice of grain properties. To compute the spectrum of a single disc model, we assume all dust resides at a single stellocentric distance, but that grains of each dust size have a temperature that depends on their size (dictated by their wavelength-dependent emission efficiency). All grain sizes between the minimum size $D_{\rm min}$ and a  maximum size of 10\,cm are summed, with weights set by the size distribution slope $q$ (where $dN/dD \propto D^{2-3q}$). Here we restrict models to $10/6<q<12/6$; below the lower bound $\sigma_{\rm tot}$ becomes dominated by large grains, and above the upper bound mass is concentrated in small grains, neither of which is thought to be the case for debris discs. The remaining  parameter is the total surface area of emitting dust $\sigma_{\rm tot}$. Given an individual disc model, the fractional luminosity can be computed by integrating the disc spectrum and dividing by the stellar luminosity. For our purposes here the benefit of this model compared to a simple modified blackbody is that the mm-wave spectral slope is restricted by reasonable assumptions about the size distribution.

To constrain the disc properties we model the optical/IR and ALMA photometry with a star + disc model. The stellar parameters of GSC\,07396-00759 are well-constrained, but the disc properties are not, with $D_{\rm min}$, $q$, and $\sigma_{\rm tot}$ spanning a wide range of parameter space. However, because our method uses MultiNest \citep{Feroz09}, the resulting distribution of disc parameters can be used to illustrate the allowed disc properties in terms of fractional luminosity and temperature, as shown in Figure \ref{fig:ft}.
While grains for any individual model have a range of temperatures, on this plot the temperature refers to the coolest grains, i.e. the grains on the scale of millimetres that have efficient emission of mm-wave radiation,  which have the same temperature as a blackbody.
While the fractional luminosity would follow a single locus for a pure blackbody model, the vertical spread of points occurs because a range of size distributions are allowed at a given dust temperature, with lower $q$ models corresponding to more blackbody-like spectra that have lower fractional luminosity, and higher $q$ models giving steeper mm-wave spectral slopes and higher fractional luminosities. Due to the differing definitions of the reported temperatures, these temperature values will be much lower than those found from a modified blackbody model and so comparisons should not be drawn between findings of the two separate modelling techniques.

Overall, Figure \ref{fig:ft} shows that the dust temperature is unlikely to be higher than 60K, and that for dust temperatures above 10K the disc fractional luminosity is greater than about $10^{-4}$.
We can also see that our models for the disc of GSC\,07396-00759 share the space occupied by other M-dwarf discs, namely TWA\,7 \citep{Bayo19}, CPD-72\,2713 \citep{Moor20,Tanner20,Norfolk21}, AU\,Mic and Fomalhaut\,C \citep{Kennedy13,Coltsmann21}, with a temperature $\sim$10-50 K and a blackbody radius $\sim$10-200\,au. However, assuming that all of the dust colocates precisely at the best-fitting ALMA radius we can narrow the probable range occupied by our models with our knowledge of the disc's observed radius which sets a limit on the blackbody temperature in the disc.

Imposing this restriction with our observed radius of 70\,au limits the model coolest grains to the blackbody temperature at 70\,au of 20\,K, i.e. on the vertical red dashed line in Figure \ref{fig:ft}. This now also constrains the fractional luminosity to at least above 2$\times 10^{-4}$, brighter than Fomalhaut\,C's disc but similar to the discs in the younger systems.

\section{Discussion}\label{sec:Discuss}
\subsection{Comparison with near-IR scattered-light observations}

\begin{figure*}\
        {\includegraphics[width=\textwidth]
        {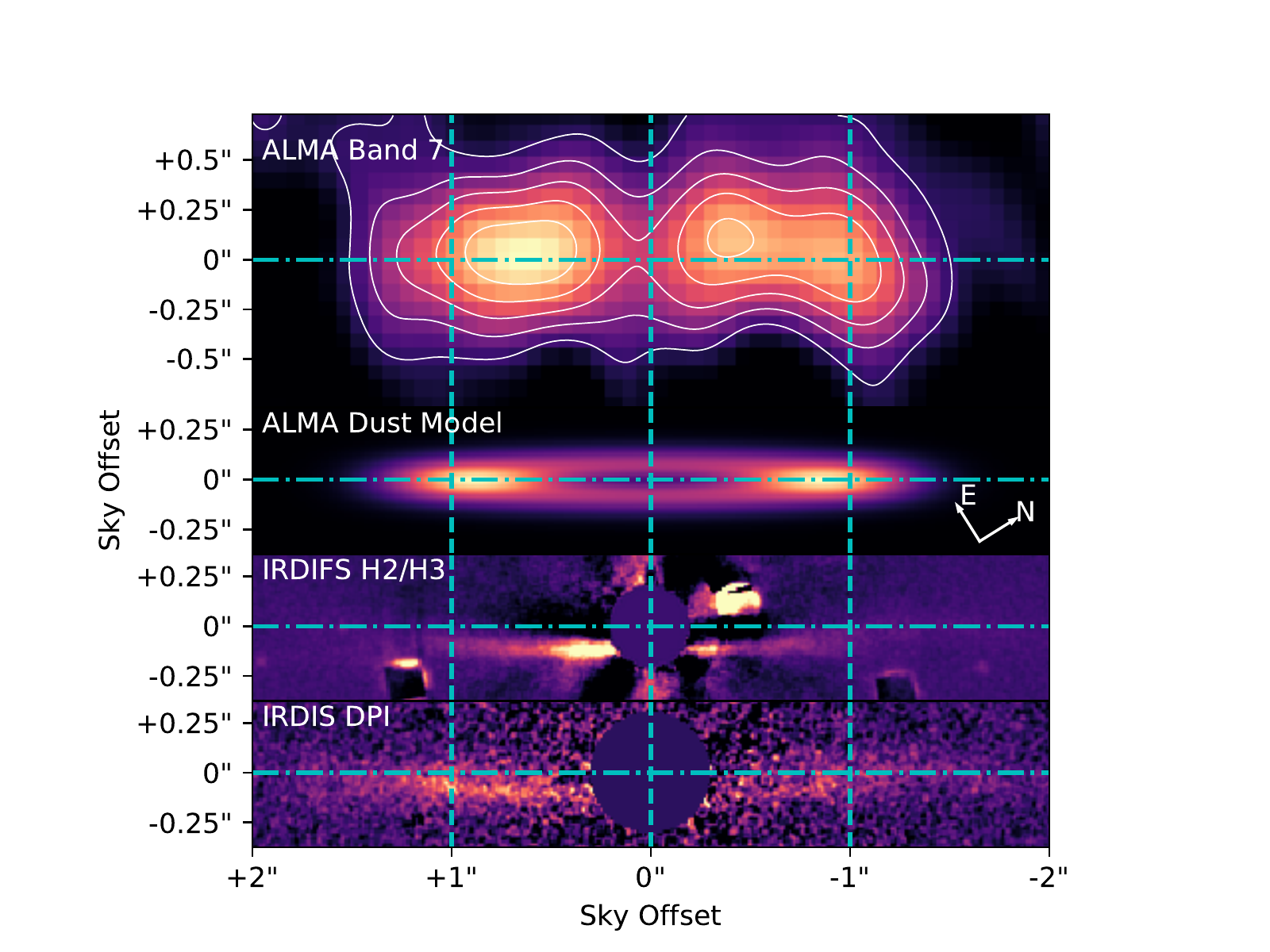}}
    \caption{\label{fig:Comparison}. Comparison of GSC\,07396-00759's disc as imaged in ALMA Band 7 with $+ 2\sigma, 4\sigma, 6\sigma, 8\sigma, 10\sigma$ contours (top, aligned with model centre at zero offset), ALMA best-fitting Gaussian torus model (second from top), SPHERE/IRDIFS H2/H3 (second from bottom) and SPHERE/IRDIS DPI (bottom). The horizontal dash-dotted lines cross through the location of the model centre for the ALMA data/model and the star for the SPHERE data, parallel to the major-axis of the disc. The central regions of the SPHERE data have been removed to account for the coronagraphic masks and high noise levels surrounding the masks. The vertical dashed lines pass through the ALMA best-fitting model radius of the disc and zero offset.}
\end{figure*}

We display a comparison of ALMA and SPHERE/IRDIS data, as well as the ALMA best-fitting Gaussian torus model in Figure \ref{fig:Comparison}, recalling that the scattering phase functions have a strong effect on how flux is distributed as a function of scattering angle for the SPHERE scattered light data.
From the data alone, the total-intensity IRDIFS flux does not appear to be more extended that the ALMA Band 7 flux, or even the ALMA dust model. The polarimetric IRDIS DPI flux however does visibly extend past $\sim1.5\arcsec$, beyond both the ALMA Band 7 flux and our underlying dust distribution model, implying that smaller micron sized dust grains are present at more distant radii than mm sized dust grains. To probe further, a comparison between the models produced for each observation is needed.

As displayed in Table \ref{tab:modelresults} the disc's position angle and inclination are consistent across all three observations' models. We however find that our best fitting model radius of 70.2\,au supports S18's model radius over A21's model radius. Even the larger radius of our double power law model with its larger error bounds is still interior to that of A21. 
As larger dust grains are a more direct tracer of the planetesimal birth ring, the ALMA derived radius is likely the most accurate, lending more weight to the total intensity scattered light model over the polarimetric scattered light model.
This would thus imply that the polarimetric model is indeed overestimating the anisotropic scattering factor $g$, as suggested by A21 themselves. That A21's fixed 70\,au radius, low $g$ model did not account for all the flux at the distant reaches of their observations indicates that the Henyey-Greenstein prescription used by A21 is probably too simple.

The ALMA double power law finds a much steeper outer slope than that of S18's model, limiting the physical dust presence inferred to less than radii of 1.5$\arcsec$.
This indicates that the shallow outer slope of S18's model and associated dust presence at outer radii, as well as the visually extensive flux of A21's observations, necessitate a strong radial pressure force to move small dust grains onto eccentric orbits. S18 note that their outer slope aligns well with predictions of the behaviour of small dust grains in the outer regions of debris discs under the effects of stellar radiation and wind pressures \citep{Strubbe06,Thebault08}.
For this low luminosity, but young and late-type system, stellar wind ram pressure is the most likely candidate for the pressure force, as posited in A21.
The steep outer slope also suggests the eccentricities of mm-sized grains and planetesimals are low, otherwise the outer slope would be smoother \citep{Marino21}.

\begin{figure}
        {\includegraphics[width=\columnwidth]
        {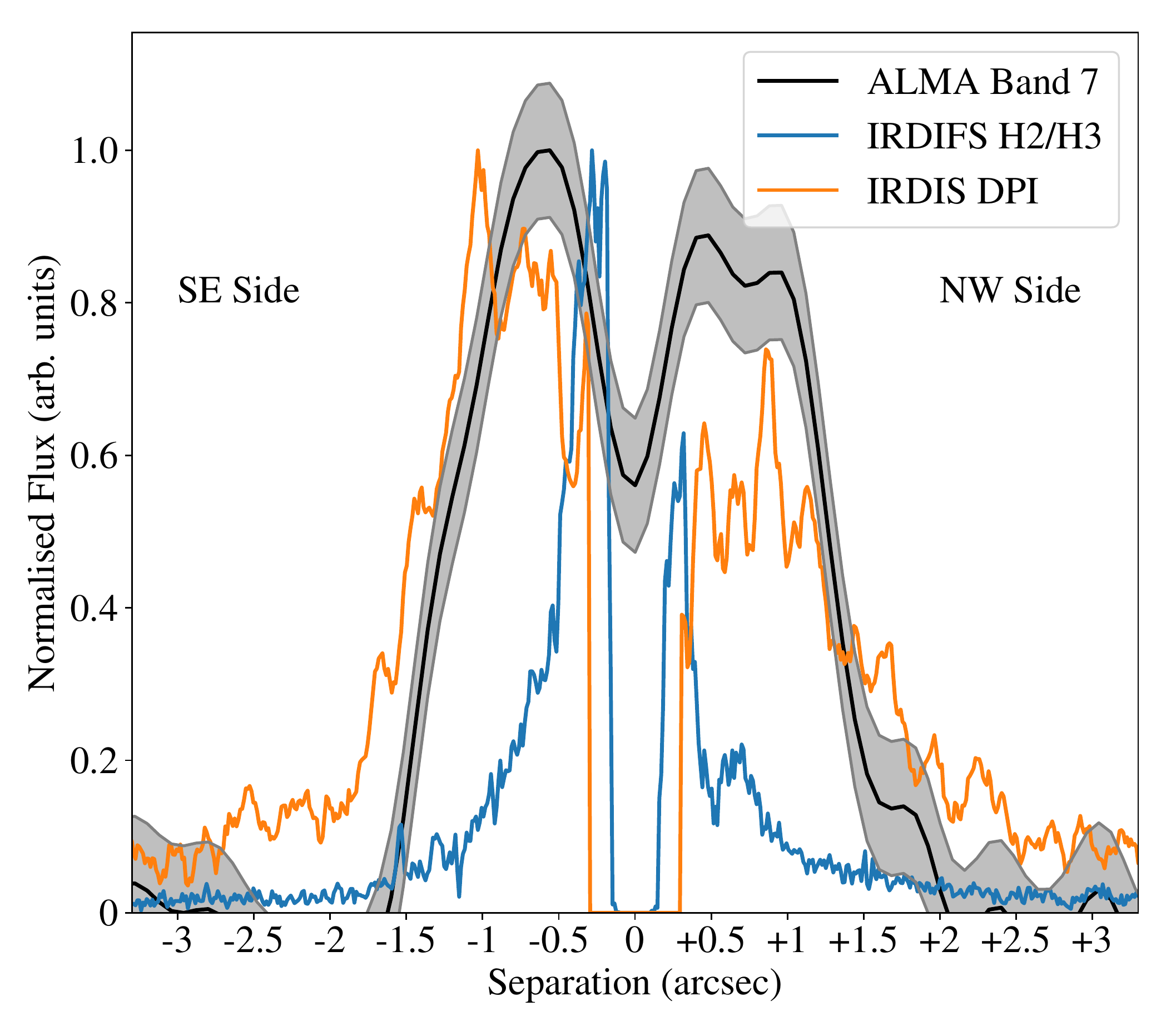}}
    \caption{\label{fig:Asymmetry}
    Comparative brightness profiles of the ALMA Band 7 and SPHERE/IRDIS total intensity and polarimetric data. For the ALMA data, the flux of the centre pixel along the disc major-axis is taken at each separation step. For the SPHERE data the peak brightness is taken from a slice of pixels at each separation step along a swathe parallel to the disc major-axis. Each profile is normalised to its brightest component. The gray swathe shows the ALMA RMS. The IRDIS DPI image has been smoothed by a uniform filter of width ten pixels. Zero separation is the best fitting model centre from \S \ref{sec:mod}.}
\end{figure}

Figure \ref{fig:Asymmetry} shows a comparison of the profiles of the three data sets, for the ALMA data derived from the flux of the centre pixel along the major-axis and for the SPHERE data derived from the brightest pixel in a slice of the disc perpendicular to the major-axis at each separation step, the differing approaches warranted due to the significantly higher resolution of the SPHERE data. The scattering phase functions will heavily dampen the visual radial extent of the disc in the scattered light data, thus Figure \ref{fig:Asymmetry} is intended for a comparison of the potential brightness asymmetries in each dataset rather than apparent radii. However, even with the effects of the scattering phase function, the IRDIS DPI profile visibly extends beyond 1.5$\arcsec$ and both the ALMA Band 7 and IRDIFS profiles. 

Both S18 and A21 observe a brightness asymmetry in the disc, with the south-east side brighter than the north-west, but by a larger factor in the June 2017 S18 total intensity data than the June 2018 A21 polarimetric data. Our April 2018 ALMA observation is consistent with there being no detected asymmetry, attributing the apparent dip on the north-west side to a noise feature. Looking at the 1$\sigma$ bounds of the ALMA profile in Figure \ref{fig:Asymmetry}, we can place a limit on a possible sub-mm brightness asymmetry of less than a factor $\sim$1.5. It is not unfeasible that a brightness asymmetry is present in the sub-mm, but we infer that it is very unlikely to be at the same level as seen by IRDIFS and unlikely to be at the same level as seen by IRDIS DPI. The time baseline between observations is too short for dust causing an asymmetry to be removed totally from the system, which would happen on orbital timescales of $\sim$750 years, and so the asymmetry must be enhanced in small grain sizes.
Thus whatever mechanism is causing the brightness asymmetry more strongly affects the more mobile small micron sized dust grains, and is not noticeably affecting the underlying planetesimal population nor the population of mm sized dust grains. This points towards a pressure force, such as interaction with the interstellar medium \citep[e.g.][]{Maness09,Debes09} or asymmetric small dust production and/or removal such as a coronal mass ejection \citep{Osten13}.
A recent massive collision could also produce a dust asymmetry that evolves differently for dust grains with different $\beta$ values, where $\beta$ is the radial force to gravitational force ratio. In M-dwarf systems $\beta$ depends strongly on the strength of the stellar wind, which is still an unknown for GSC\,07396-00759. As $\beta$ is size dependent, this would result in differently apparent asymmetries depending on the grain size probed by the observation \citep{Jackson14,Kral15}.

A warp in the north-west of the disc is observed in both scattered light data sets. In Figure \ref{fig:Comparison} slight evidence for a warp in the ALMA data may be visually identified in the north-west, but it is difficult to extricate this from the larger noise features overlapping the north-west of the disc as seen in Figure \ref{fig:res}. The observation most likely does not possess the necessary resolution to uncover a warp if one exists in the sub-mm grain and parent planetesimal distributions. Higher resolution sub-mm follow up observations would allow this to be investigated.

\subsection{Comparison with other M-dwarf discs}

\begin{figure}
        {\includegraphics[width=\columnwidth]
        {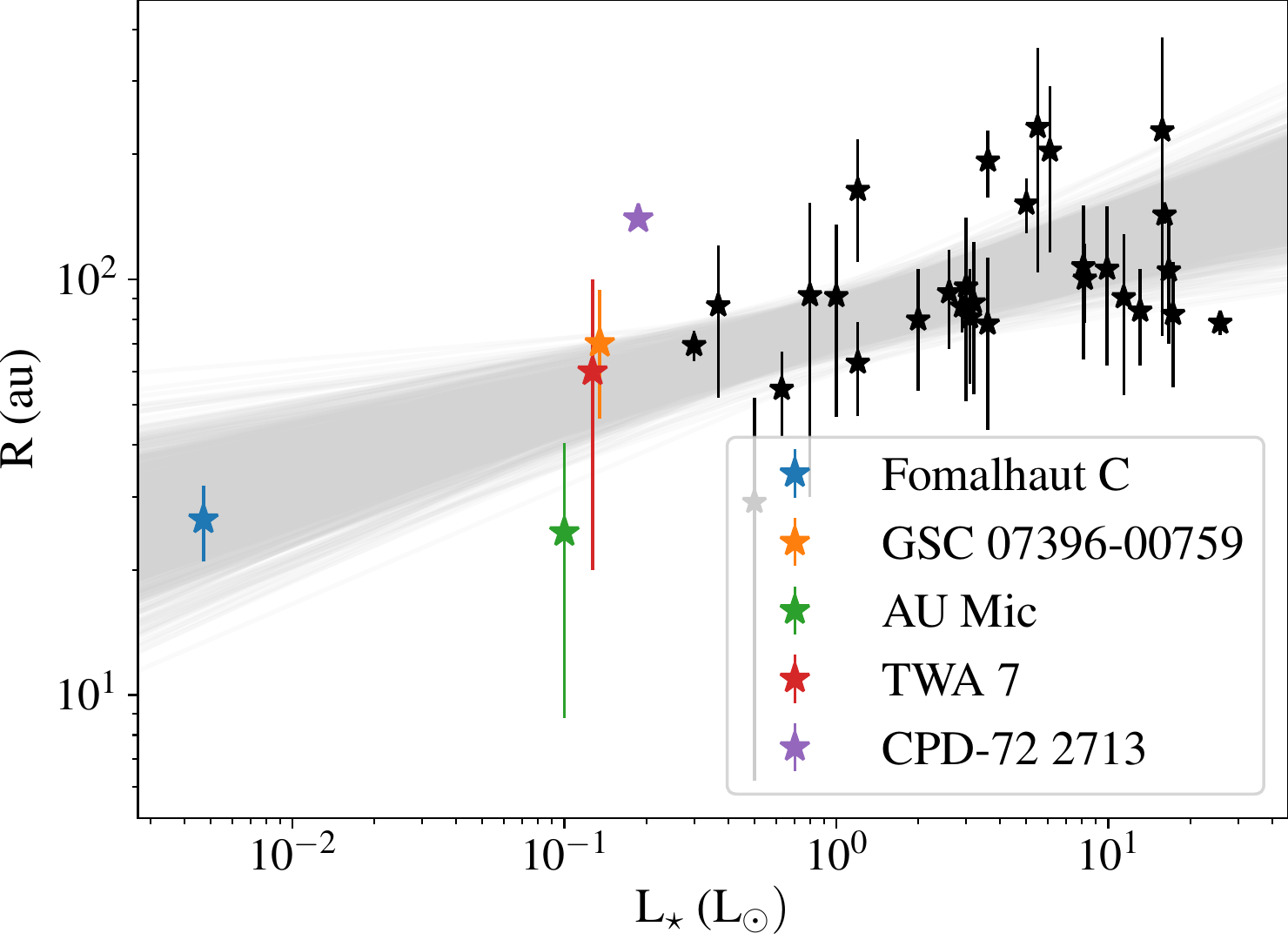}}
    \caption{\label{fig:LR} mm-wave resolved debris disc radii plotted against host stellar luminosity. Error bars represent disc FWHM or upper limits. The five latest type stellar hosts are highlighted in colour, CPD-72\,2713 is plotted without a width  as a fixed width of 0.2R was assumed to facilitate fitting a radius \citep{Moor20}. Transparent grey lines show a sample of 1000 power laws from the parameter distributions of \citet{Matra18}.}
\end{figure}

GSC\,07396-00759's disc's radius of 70\,au is about double that of AU\,Mic's disc's radius of $\sim$24-40\,au \citep{Daley19}, and we find GSC\,07396-00759's disc to be of intermediary width when the widths are presented as ratios to their radii: a FWHM of 50\,au for GSC\,07396-00759 gives a ratios of 0.7 and \citet{Marino21} finds a ratio of 0.97 for AU\,Mic, in comparison to the M-dwarf host Fomalhaut\,C's thinner disc with radius 26.4\,au, FHWM upper limit of 11\,au and ratio upper limit of 0.42. 
GSC\,07396-00759's disc's radius is comparatively similar to the $\sim$60\,au radius of the face-on debris disc around the young \citep[7.5$\pm$0.7 Myr old,][]{Ducourant14} M2Ve star TWA\,7 derived from marginally resolved ALMA observations \citep{Bayo19,Matra19}.
TWA\,7 has also been shown to possess considerable structure when viewed in scattered light \citep{Ren21,Olofsson18}. The asymmetry of GSC\,07396-00759 could be similar to that of TWA\,7, if TWA\,7 were viewed edge-on, which is apparent in scattered light but only marginally identified in the sub-mm \citep{Ren21}. CPD-72\,2713's disc remains unusually large for its type with its radius of 140\,au \citep{Moor20}, twice that GSC\,07396-00759's.

To visualise these comparisons we place GSC\,07396-00759 on the radius-luminosity plot presented in \citet{Matra18}, with the addition of the sample presented in \citet{Sepulveda19}, Fomalhaut\,C \citep{Coltsmann21} and CPD-72\,2713 \citep{Moor20}. CPD-72\,2713 is presented without a disc width as a fixed FWHM of 0.2R was used to reduce degeneracy while fitting for a radius in the marginally resolved observation. The sample of mm-wave resolved discs at M0$\sim$M2 is growing and is appearing to remain largely consistent both within the subset and with the greater planetary belt demographic, both in terms of the average of the radii across the sample and the breadth of the spread of their radii.

The growing sample of these discs that are resolved in both the sub-mm and scattered light will also help to elucidate the mechanisms of stellar wind forces in this regime where the low luminosity of the host star is insufficient to remove dust grains via radiation pressure and stellar wind forces become dominant, as is the case for GSC\,07396-00759 \citep{sissa18,Adam21}.

\subsection{CO non-detection}

We searched for evidence of volatiles released by planetesimal collisions via the CO gas J$=$3-$2$ emission line, as per our ALMA spectral setup. We subtracted the continuum emission and visually inspected both the dirty cube and the moment-0 map collapsed over the range of velocities where emission would be expected, finding no immediate signal.

To increase our sensitivity to a small amount of CO we use the spectro-spatial filtering technique first described in \citet{Matra15}, assuming that CO would be co-located with the dust. 
This method shifts pixels in the spectral cube based on the expected Keplerian orbital velocity at their location into a single channel to enhance signal; we have assumed a stellar mass of 0.62\,M$_\odot$ \citep{Adam21}.
However, the edge-on disc, low stellar host mass, large disc radius and low spectral resolution all limit the effectiveness of the technique.  
Our spatial filter is taken from our Gaussian torus model from \S \ref{subsec:Torus}, masking all pixels not co-located with model continuum emission of at least 10\% the maximum emission. 

\begin{figure}
        {\includegraphics[width=\columnwidth]
        {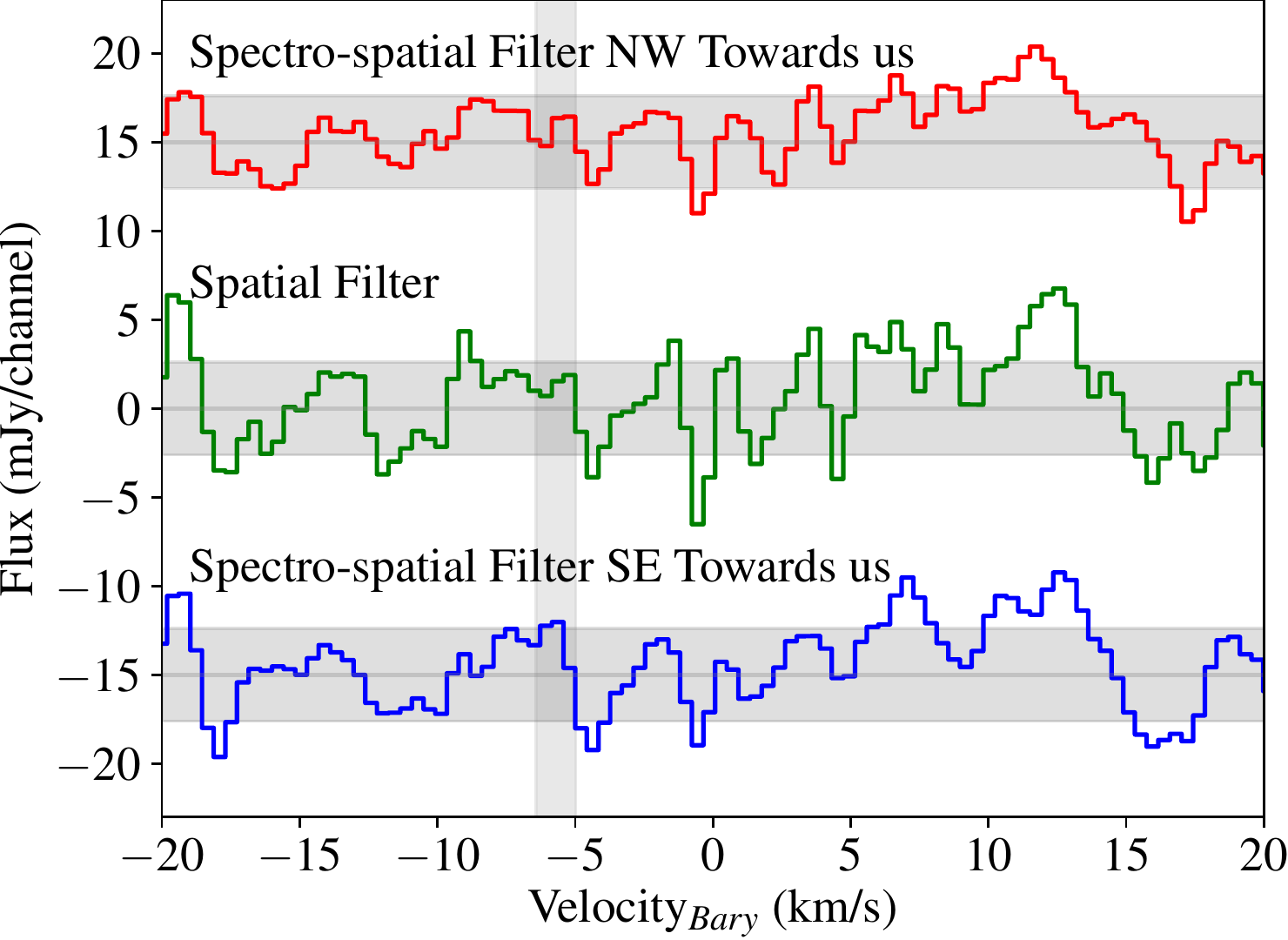}}
        \caption{\label{fig:CO} Spatially and spectro-spatially filtered CO J=3-2 spectra for the debris disc around GSC\,07396-00759.
        The 1$\sigma$ uncertainty of the spectrum is measured over a larger range of velocities and is denoted by the horizontal shaded regions. The expected centre of the signal at the -5.7$\pm 0.8$\,km\,s$^{-1}$ stellar radial velocity is denoted by the vertical shaded region.
        }
\end{figure}

In Figure \ref{fig:CO} we display the spectra corresponding to the spatial filter alone, and spectro-spatial filters with either assumption of the north-west section of the disc rotating towards us, or the south-east section rotating towards us. 
We do not detect any trace of CO gas and instead find a 3$\sigma$ upper limit on the CO flux of 30\,mJy\,km\,s$^{-1}$, calculated from the RMS in combination with ALMA's 10\% flux calibration uncertainty and the effect of the correlation of adjacent channels. 

We can compare this detection limit to the detection of CO\,J$=$3$-$2 emission in the disc of the similar luminosity M-dwarf TWA\,7 \citep{Matra19}. At a distance of 34\,pc an integrated flux of 91$\pm$20\,mJy\,km\,s$^{-1}$ was measured for TWA\,7; scaling our limit for GSC\,07396-00759 to this distance gives a limit at 34\,pc of 132\,mJy\,km\,s$^{-1}$. This means that our observations would not have detected a TWA\,7 analogue, i.e. if GSC\,07396-00759 shared the same collisional mass loss rate, photodissociation timescale, excitation environment and CO mass as TWA\,7, but our limit would have been close to the underlying CO flux.

 \citet{Matra19} derive an already large CO ice fraction of $\geq70\%$ for TWA\,7, thus we can conclude that our non-detection likely does not constrain the presence of CO in the system or its ice fraction in planetesimals in a meaningful way; but we can at least say that the disc is not gas-rich, solidifying its status as an evolved debris disc and not a primordial disc like that of GSC\,07396-00759's companion system V4046\,Sgr.

\section{Conclusions}

We have resolved the debris disc around the M1V star GSC\,07396-00759 in mm-wave thermal emission for the first time, making it one of only a small handful of M-dwarfs with resolved debris discs, and one of only two both fully resolved in scattered light and thermal emission along with AU\,Mic. We model the geometry of the underlying dust distribution, and inferred birth ring of planetesimals, as revealed by ALMA, well constraining the disc radius to $70.2^{+4.1}_{-4.7}$\,au and sub-millimetre flux to $1.84^{+0.22}_{-0.21}$\,mJy. We trial a simple Gaussian disc model as well as a double power law model to investigate the radial extent of the mm dust grains and find the Gaussian model to be the more appropriate fit. We do not find the disc centre to be significantly offset from the stellar location and so place an upper limit on any underlying disc eccentricity along the major-axis. We also do not detect the presence of CO gas within the system, further distancing the evolutionary states of this debris disc and the primordial disc around the associated star V4046\,Sgr.

We compare our sub-millimetre findings with previous scattered light observations, in both total intensity \citep{sissa18} and polarimetry \citep{Adam21}. Our double power law model has a significantly steeper outer slope in the dust distribution that the total intensity model, i.e. \citet{sissa18} infer micron sized dust grains to be present at much larger radii than we infer mm size dust grains to be present at. \citet{sissa18}'s model thus requires a radial pressure force predominantly affecting smaller dust grains, most likely the action of stellar wind in this low host star luminosity system.
Our sub-millimetre radius measurement is a stronger tracer of the underlying planetesimal belt and so confirms the radius measurement made by \citet{sissa18} over \citet{Adam21}, implying that complex behaviour of the polarised scattering phase function was responsible for the large radius measurement made by the latter. 

We do not significantly detect in the sub-millimetre the brightness asymmetry apparent in both the scattered light observations. This implies that the physical mechanism behind the asymmetry is a pressure force acting on smaller dust grains or related to asymmetric production/removal of small dust. However, higher signal to noise sub-millimetre observations could still reveal an asymmetry in the mm dust grains.
Our ALMA observations are also not of significantly high resolution to identify any warps in this disc, as also observed in scattered light.

We do not have enough measurements of the disc flux in the far-infrared/sub-millimetre to constrain an SED for the disc. However, we do explore the possible disc fractional luminosity/representative dust temperature parameter space to identify that the disc around GSC\,07396-00759 is likely to possess a greater fractional luminosity than the disc around Fomalhaut\,C, and could have a similar or even greater fractional luminosity than the discs around low mass stars AU\,Mic, CPD-72\,2713 and TWA\,7. The radius of GSC\,07396-00759's disc, almost thrice that of AU\,Mic's disc but similar to that of TWA\,7's, places it in good agreement with the proposed radius-luminosity relationship proposed by \citet{Matra18}, and the disc width is moderate among the greater population of mm-wave resolved debris discs.

As an edge-on M-dwarf debris disc well resolved both in the sub-millimetre and in scattered light, and with dust features apparent in the scattered light that are not present in the sub-millimetre, GSC\,07396-00759 is an excellent candidate for follow-up observations, to investigate low host luminosity stellar wind dominated discs and the source of the system's own unique features as well as to finally provide a true coeval disc to compare the discoveries from its twin AU\,Mic with.

\section*{Acknowledgements}
PFCC is supported by the University of Warwick.
GMK is supported by the Royal Society as a Royal Society University Research Fellow. 
The research of CA is supported by the Comité Mixto ESO-Chile and the VRIIP/DGI at University of Antofagasta.
J.\,O. acknowledges support by ANID, -- Millennium Science Initiative Program -- NCN19\_171, from the Universidad de Valpara\'iso, and from Fondecyt (grant 1180395).
LM acknowledges funding from the European Union's Horizon 2020 research and innovation programme under the Marie Sklodowska-Curie grant agreement No 101031685.
SM is supported by a Junior Research Fellowship from Jesus College, University of Cambridge.
This paper makes use of the following ALMA data: ADS/JAO.ALMA\#2017.1.01583.S. ALMA is a partnership of ESO (representing its member states), NSF (USA) and NINS (Japan), together with NRC (Canada), MOST and ASIAA (Taiwan), and KASI (Republic of Korea), in cooperation with the Republic of Chile. The Joint ALMA Observatory is operated by ESO, AUI/NRAO and NAOJ.
This work has made use of data from the European Space Agency (ESA) mission {\it Gaia} (\url{https://www.cosmos.esa.int/gaia}), processed by the {\it Gaia} Data Processing and Analysis Consortium (DPAC, \url{https://www.cosmos.esa.int/web/gaia/dpac/consortium}). Funding for the DPAC has been provided by national institutions, in particular the institutions participating in the {\it Gaia} Multilateral Agreement.
\section*{Data Availability}

The data underlying this article are available in http://almascience.nrao.edu/aq/, and can be accessed with ALMA project ID: 2017.1.01583.S.
Data used to produce the VLT/SPHERE images and profiles in Figures \ref{fig:Comparison} and \ref{fig:Asymmetry} were reproduced from \citet{sissa18} and \citet{Adam21}.



\bibliographystyle{mnras}
\bibliography{GSC} 



\appendix

\section{Modelling posteriors}

\begin{figure*}
        {\includegraphics[width=\textwidth]
        {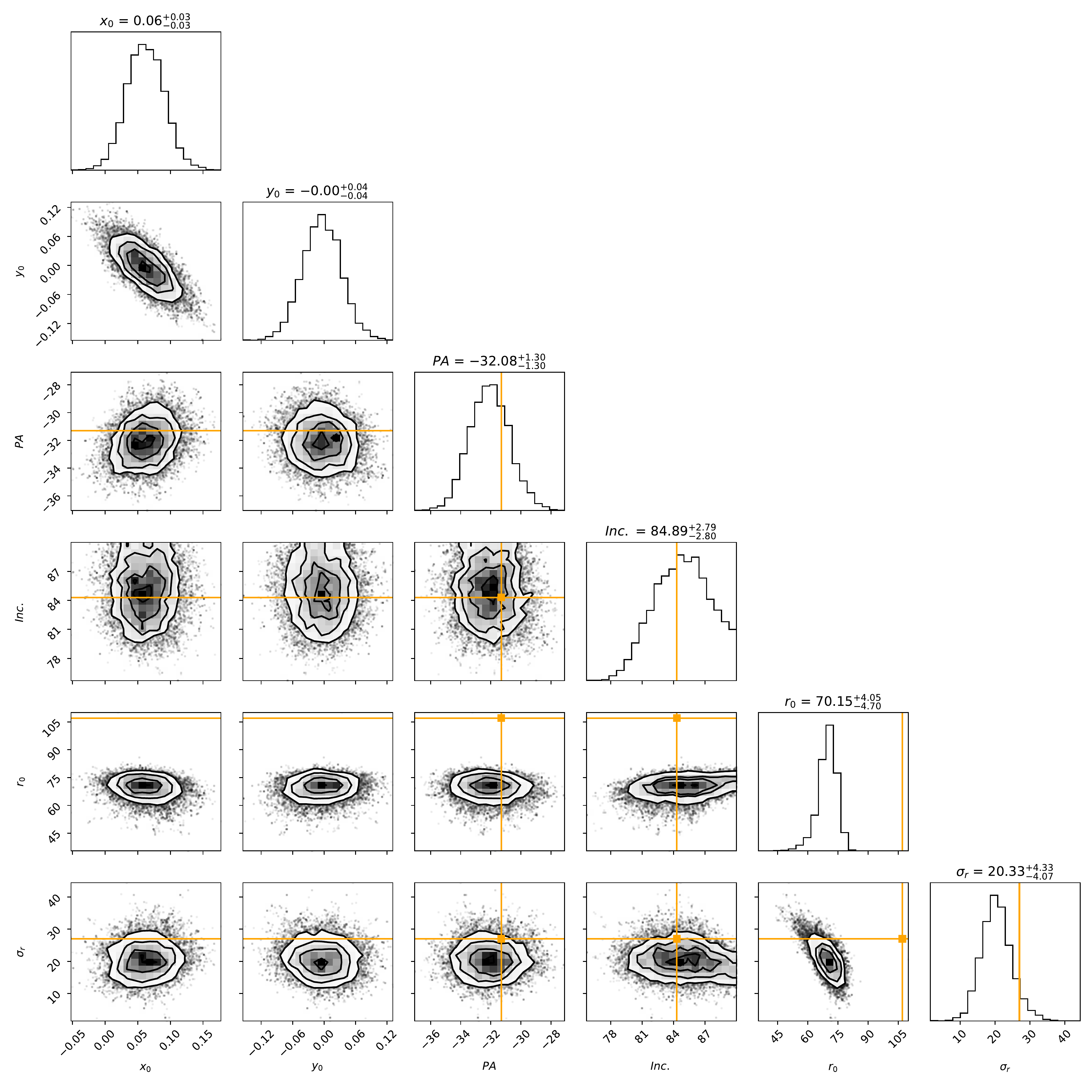}}
        \caption{\label{fig:CornerGau} Posterior distributions of parameters from MCMC fitting of the Gaussian disc model. Where model parameters are shared, the median parameters of \citet{Adam21} are overplotted in orange for comparison.
        }
\end{figure*}

\begin{figure*}
        {\includegraphics[width=\textwidth]
        {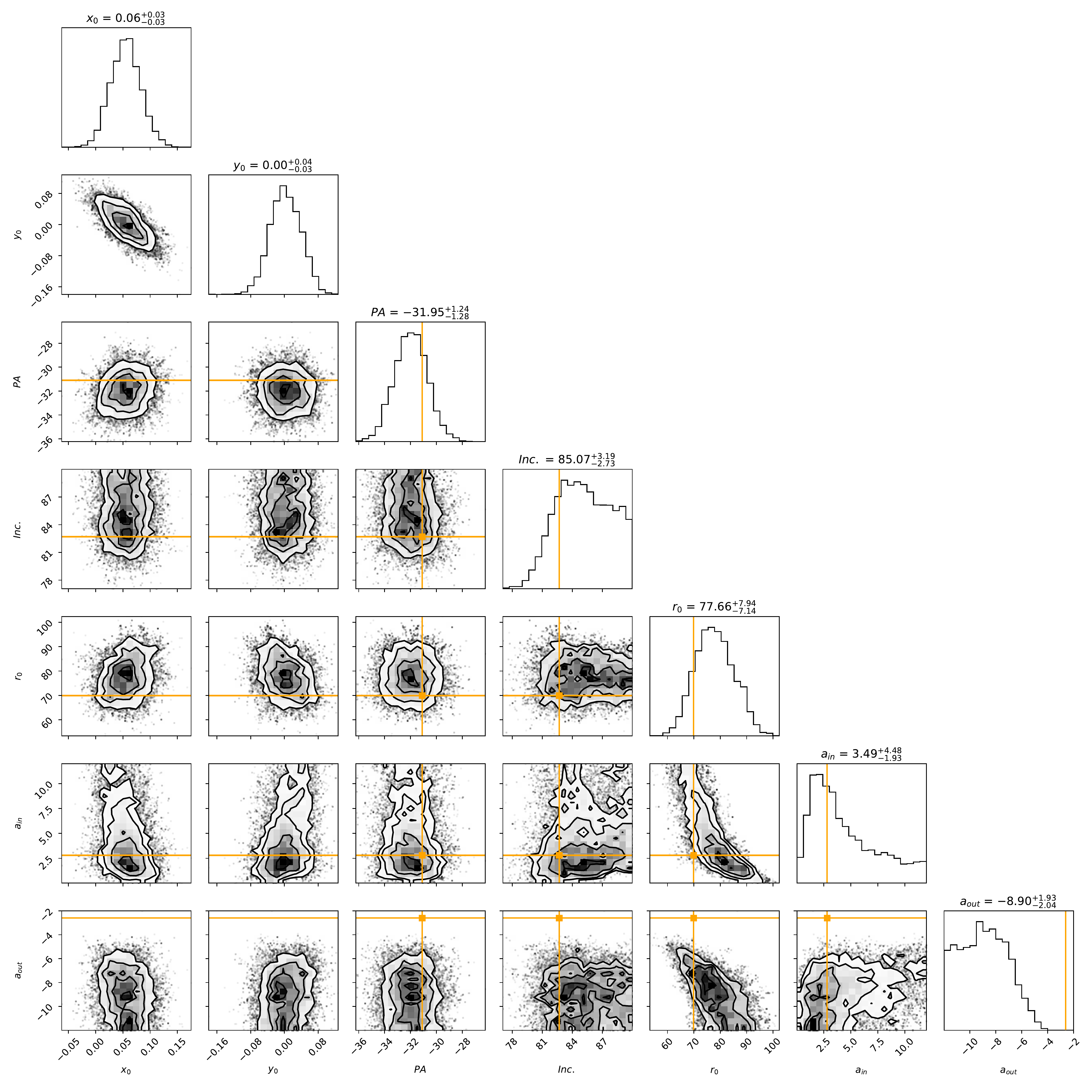}}
        \caption{\label{fig:CornerPow} Posterior distributions of parameters from MCMC fitting of the double power law disc model. Where model parameters are shared, the median parameters of \citet{sissa18} are overplotted in orange for comparison.
        }
\end{figure*}

Figure \ref{fig:CornerGau} and figure \ref{fig:CornerPow} show the posterior distributions of select parameters from the MCMC fitting of the Gaussian and double power law models respectively, where parameters are shared between our models and S18 or A21's we have plotted their median parameters for comparison. As displayed in Figure \ref{fig:CornerGau} the Gaussian model only shows a large degeneracy in the $x$ and $y$ offsets and a slight degeneracy between $r_0$ and $\sigma_r$. As displayed in Figure \ref{fig:CornerPow} the double power law model as well as the $x$ and $y$ offset degeneracy there are significant degeneracies between $r_0$, $a_{\rm{in}}$ and $a_{\rm{out}}$. 

\bsp	
\label{lastpage}
\end{document}